%% file: WSDM2018.tex
\def\year{2018}\relax

\documentclass[letterpaper]{article} 
\usepackage{aaai18}  
\usepackage{times}  
\usepackage{helvet}  
\usepackage{courier}  
\usepackage{url}  
\usepackage{graphicx}  
\input{sections/frontmatter}
\author{
	Pushkal Agarwal$^1$,  Nishanth Sastry$^1$,  Edward Wood$^2$ \\ 
    $^1$King's College London, $^2$House of Commons Research Library
}
\begin{document}
%
\title{Discovery of the Content and Engagement with the Content}
\maketitle

\input{sections/introduction}
\input{sections/relatedWork}
\input{sections/framework}

\input{sections/application}
 \input{sections/conclusion}

\input{sections/figdump}
{
	\balance
    {
		\bibliographystyle{ACM-Reference-Format}
		\bibliography{bib/references} 
	}
}

\input{sections/story}
\end{document}

%% file: sections/frontmatter.tex
\usepackage{url}
\usepackage{graphicx,epsfig}
\usepackage{subfig}
\usepackage{graphics,dcolumn,bm,fleqn,float}
\usepackage{amssymb,amsmath,multirow,rotate,color}
\usepackage{epstopdf}
\usepackage{soul}
\usepackage{listings}
\usepackage{xcolor}
\usepackage{booktabs}
\usepackage{natbib}
\usepackage[T1]{fontenc}
\usepackage{xspace}
\usepackage[flushleft]{threeparttable}
\usepackage{booktabs}
\usepackage{balance}
\usepackage[inline]{enumitem} 

\newlist{RQ}{enumerate}{1}
\setlist[RQ]{label=\textbf{RQ}-\arabic*,resume,leftmargin=*}

\renewcommand{\footnotesize}{\scriptsize}

\newcommand\reffig[1]{Fig.~\ref{#1}}

\def\eg{\emph{e.g.,}\xspace}



%% file: sections/introduction.tex
\section{Introduction}
\label{sec:intro}


For much of the 20th century, British politicians bemoaned the decline in reporting of Parliament by the press. The UK Parliament's \emph{Modernisation Committee} noted ruefully in 2004 that ``the House can no longer expect to receive a certain amount of media coverage as of right.  Parliamentary proceedings must now compete with other potential news stories for coverage''\footnote{26.5.04, para 108, \url{https://publications.parliament.uk/pa/cm200304/cmselect/cmmodern/368/36809.htm}}. One obvious strategy to counter this trend is for Parliament to provide direct coverage of debates and committee meetings, allowing people to get news straight from the horse's mouth. This window into the UK Parliament's activities is provided in three forms: \emph{transcripts, video and social media}. The first is \emph{Hansard}, which is a near verbatim transcript of the Parliamentary debates and committee sessions. \emph{Hansard}, which took its name from the printer of the early transcripts, is one of the oldest such series in the World: regular reporting of the UK Parliament has been in operation since the early 19th century. Access to \emph{Hansard} was increased dramatically when it was published online from 1996.  In the second half of the 20th century, Parliament allowed broadcasters to transmit radio and eventually television coverage of debates and meetings of select committees.  More recently, in an effort to further improve transparency and citizen engagement, the UK Parliament started publishing videos of these debates and meetings itself, and tweeting details of debates as they happened. 

In this paper, we attempt to characterise how people engage with video data of Parliamentary debates by using more than two years of Google Analytics data around these videos\footnote{Data provided by the Video Analytics team of UK Parliament}. We analyse the patterns of engagement -- how do they land on a particular video? How do they hear about this video, i.e., what is the (HTTP) referrer website that led to the user clicking on the video? Once a user lands on a video, how do they engage with it? For how long is the video played? What is the next destination? etc. 

Answering these questions is an important first step towards understanding why and how people use Parliamentary videos, and therefore, how the video delivery platform should be adapted and personalised for the needs of the citizens of the country. Taking inspiration from \cite{an2017personas}, we employ Non-Negative Matrix Factorization (NMF)~\citep{lee1999learning} on  the video views matrix to identify different archetypes of users, and identify 5 archetypes. A deeper examination of the archetypes we find reveals that they are primarily distinguished by how they land on the video page: Search (i.e., through a search engine), Referral (i.e., from other Parliamentary websites), Direct (i.e., through a direct link, which is embedded on another website), Social (i.e., through a social platform such as Facebook or Twitter) and Others. 

Interestingly, these different archetypes appear to have different levels of engagement with the Parliamentary videos: Those that land on a video from a social website tend to watch the video for a shorter duration than those that land from other sources such as referrals or direct links.  When a larger number of clusters is generated, further differences can be amongst users who land from a social platform, with Twitter forming a separate cluster as compared with Facebook and Reddit as more number of MPs are active on Twitter ~\citep{agarwal2019tweeting}. Also, the different archetypes have different viewing interests which could be used for improving personalization for the end user. As an example, during the period of our analysis, social media users were mostly interested in debates relating to the ongoing separation of the United Kingdom from the EU (also known as ``Brexit'').

In the rest of this extended abstract, we first describe the dataset, and then our initial results. Finally we sketch implications for personalisation. 

%% file: sections/framework.tex
\section{Dataset and Algorithm} \label{sec:dataset}
UK parliament session video are broadcasted live at \emph{parliamentlive.tv} for all to view. Subsequent to the live broadcast, an \emph{archive} of the video is also saved for on-demand viewing at any later date. Citizens from UK and other countries can view these live and archived sessions, check a calendar for the schedule of sessions and so on. Each session also gets its own URL, and users can share the URLs on OSMs (Online Social Media). 

Our analysis is based on the \emph{viewing logs} and referral source of landing users of these session videos. The details of \emph{Google Analytics} data which we explored in the study are in Table \ref{table:data} 

\begin{table}[h]
\centering
 	\begin{tabular}{|p{3cm}|p{4cm}| } 
		\hline
      	Duration& 23-02-2015 to 05-02-2017\\
      \hline
 		Total Views& 5.96 Million\\
      \hline
 		Total Viewers& 4.1 Million\\
      \hline
 		Referral Groups& 1,056\\
      \hline
 		Total Videos& 17,012\\
      \hline
	 	Total Video types& 6,251\\
      \hline
	\end{tabular}
    \caption{Dataset Details}
    \label{table:data}
    \vspace{-0.5cm}
\end{table}


One of the main reasons for running these video broadcasts and other online parliamentary services is to promote user engagement. However, ``the public'' does not seem to have an overwhelming interest in watching parliamentary debates: the BBC Parliamentary channel had the least views in analysis of six million users of iPlayer \citep{Nencioni:2016:SEG:3013746.3013782}. So, one question to address here is how can we make it more engaging?

Google Analytics' dashboard of the aforementioned data claims that 42\% users landed from a \emph{Direct Search} \eg Video link on homepage, 22\% from other parliament websites' \emph{Referral}, 20\% from \emph{Social} and 14.7\% from organic \emph{Search}.  For parliament websites' referral, the  average session duration is 3 minute 11 seconds while for social it is 1 minute 36 seconds. The highest average duration for any class is 4 minute 11 second, for organic search. This suggests that there are strong differences in consumption behavior. Also, a thing to note here is that direct  referral and search are somewhat offline recommendation where users themselves browse for and start playing a video. In these cases users discover the content with no online influence or personalization. For users landing from social platform, their personalized views from their social news feeds affect which videos are visible to them. 

To understand more about these different patterns in viewing behavior we grouped the raw data from Google Analytics and then extracted features of users' components. To achieve the first part we created viewership matrix $V$, where $V_{ij}$ represents number of views coming from an $i^{th}$ referral group (\eg Google, Twitter, E-mails and so forth) for all $j^{th}$ video type (\eg all house of commons videos) for entire duration of the data.

After computing users' engagement with respective content as matrix $V$, we use Non-Negative Matrix Factorization(NMF) to obtain $W$ and $H$ such that $V \approx W H$. NMF is a state of the art technique to separate features from groups based on a rank of the matrix supplied as $p$. \citep{an2017personas} used this to propose a model of Automatic Persona Generation (APG). Since Google Analytics gives \emph{five} types of online user groups (Search, Referral, Direct, Other, Social) we use $p=5$. Complete NMF is in equation \ref{eq:APG}, where Error matrix $E$ balances the approximation on RHS.
\begin{equation}\label{eq:APG}
V_{gXc}=W_{g \times p} X H_{p \times c}+ E_{g\times c}
\end{equation} 
The application of NMF on our data and features of users' group is in the next Application section.




%% file: sections/application.tex
\section{Application} \label{sec:application}

\begin{figure}
   \vspace{-0.3cm}
	\centering
	\includegraphics[trim={0 0 0 3cm},width=\columnwidth]{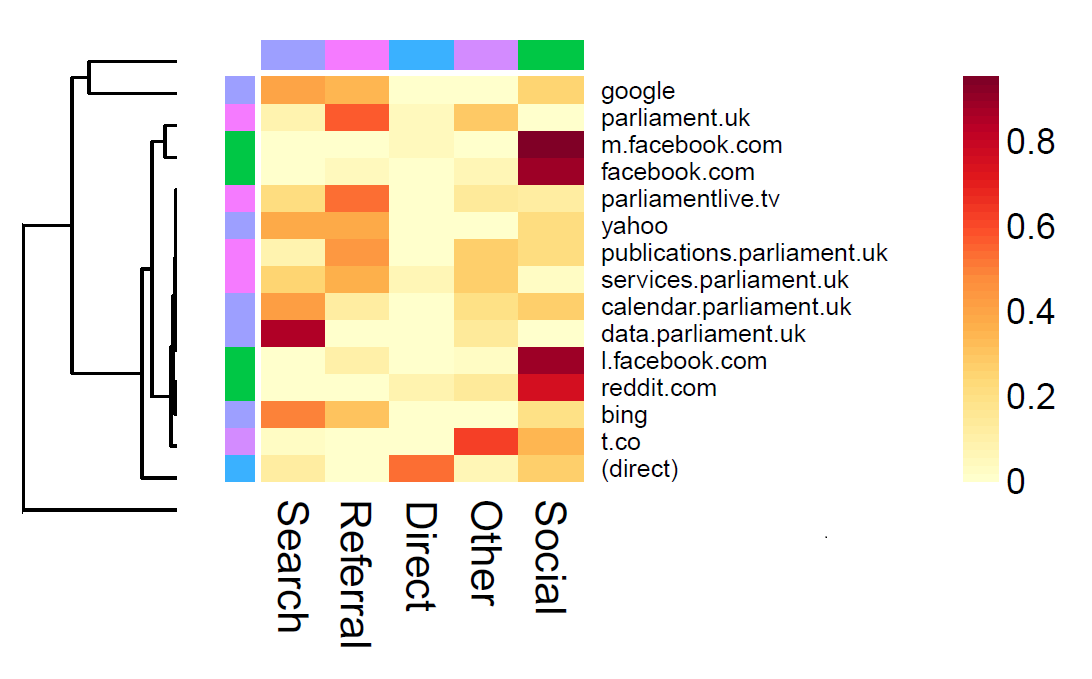}
	\caption{Top fifteen referral sources (in rows) with heat of user groups (in columns).}
	\label{fig:component}
    \vspace{-0.6cm}
\end{figure}

Using the two factors $W$ and $H$ of $V$ (from equation \ref{eq:APG}), here we check discovery and engagement with the content. Matrix $W$ represents interaction of referral source with that user groups (here five user groups as p=5). Top 15 referrals are plotted as a heatmap in \reffig{fig:component}. Higher the heat of a component, the higher the interaction between the referral source and the particular user group. This grouping based on referral source through light on view pattern and conversion of users. The Figure clearly shows that different user groups have different video preference and interests. Even Twitter (here t.co) have different user referrals from corresponding social. On the one hand online (social) users could be influenced with the personalization of their online feeds while on the other hand offline (organic search) and recommended (direct and referral) users are mostly those who hear about the event, see the event going on parliamentary website or calenders. Given their prior knowledge of the existence of the debate and their interest in finding the video, these users are more committed  and stay for a longer duration.

Another factor of matrix $V$ is $H$ which represents the engagement of user groups with the video content. Table~\ref{table:APG++} shows the type of top video on which the groups land. Interestingly, social is consuming more of Brexit-related videos, which highlights the interest of the general public and especially the social networkers in this topic.


\begin{table}[h]
\vspace{-0.1cm}
\centering
 	\begin{tabular}{|p{1.5cm}|p{5.5cm}| } 
		\hline
		Component& Top Video Type\\
        \hline
         Search&business, innovation and skills committee and work and pensions committee\\
         Referral& house of commons\\
         Direct&  westminster hall\\
		 Other&house of lords\\
         Social&exiting the european union committee\\
         \hline
	\end{tabular}
    \caption{Preferred video for user groups from matrix $H$.}
		\label{table:APG++}
\vspace{-0.6cm}
\end{table}



%% file: sections/conclusion.tex
\section{Prospects for personalisation} \label{sec:conclusion}
These preliminary results present several opportunities in the context of increasing user engagement. We identified that there are different kinds of users who consume the Parliamentary videos in different ways. It may help engagement with the Parliament if different videos were highlighted differently to these different archetypes of users. For instance, it may be useful to promote Brexit-like videos to users  on social platforms. Other kinds of debates may be more interesting to other kinds of users At the same time, different modes of delivery may be employed for different user archetypes -- for users arriving from social media platforms, since their attention span is limited, a ``summary'' or ``highlights'' version of a debate, similar to highlights shown in sports matches, may be a useful addition.